\documentclass{ragtime} % NO optional arguments are required
\usepackage{physics}
\title[Magnetized collapsars and black hole spin-up]%
      {Accretion induced black hole spin up in magnetized collapsars}

\author[A. Janiuk, \& D. Kr{\'o}l]   % the last one with `and' without a comma.
       {Agnieszka Janiuk\at[,]{1,a} % Makes referencing superscript `1'
         \& Dominika {\L.} Kr{\'o}l\at[]{2}\\\\   % ref. superscr. `1,a', but empty [] suppresses comma
       \ins{1}Center for Theoretical Physics, Polish Academy of Sciences, Al. Lotnikow 32/46,\splitins[1] %This is how to break an affiliation into two lines
       P-02-668 Warsaw, Poland\\% Termination of the affiliation
       \ins{2}Jagiellonian University Astronomical Observatory,  ul. Orla 171,\splitins[1]
       P-30-244 Krakow, Poland\\% 
       \ins{a}\Email{E-mail:agnes@cft.edu.pl}} % This is how to present E-mail.
\begin{document}     
\newcommand{\apj}{ApJ}
\newcommand{\apjs}{ApJS}
\newcommand{\apjl}{ApJL}
\newcommand{\pasp}{PASP}
\newcommand{\mnras}{MNRAS}
\newcommand{\aj}{AJ}
\newcommand{\nat}{Nature}
\newcommand{\nar}{NewAR}
\newcommand{\na}{NewA}
\newcommand{\aap}{A\&A}
\newcommand{\aaps}{A\&AS}
\newcommand{\icarus}{Icar}
\newcommand{\araa}{ARA\&A}
\newcommand{\aapr}{A\&ARv}
\newcommand{\aplett}{Astrophysical Letters}
\newcommand{\prd}{Phys. Rev. D}

\begin{abstract}

  Black holes are the final stage of gravitational collapse process, and due to the
  cosmic censorhip conjecture, they are created inevitably if a trapped surface has
  formed in the space-time. The solutions of Schwarzschild and Kerr are describing the
  spacetime metric for the gravitational field of a spherically symmetric, or
  rotating black hole. Astrophysically, the rotating black holes of stellar mass are
  end products of stellar evolution, when the progenitor star was massive enough and
  possessed a substantial amount of angular momentum. They can be discovered when leaving
  behind a luminous transient in a form of gamma ray burst, which is
  followed by an afterglow emission at lower energies and associated with the
  emerging supernova-like spectra
  that trace the chemical composition of expanding shells from the explosion.
  The gravitational binding energy of the massive progenitor star is released in the supernova explosion, while the extraction of rotational energy of the newly formed black hole
  drives the gamma ray burst. In the latter, magnetic fields are the agent driving
  the process.

  In this article, we study the gravitational collapse and formation of the Kerr black hole from the rotating progenitor star. We follow the evolution of black hole spin, coupled with its increasing mass. We study the effect of different level of rotation endowed in the progenitor's envelope, and we out some constraints on the final black hole parameters.

  Our method is based on semi-analytical computations that involve stellar-evolution models of different progenitors. We also follow numerically the black hole evolution
  and spacetime metric changes during the collapse, via General Relativistic MHD modeling.
  
\end{abstract}


\section{Introduction}\label{introduction}%%%%%%%%%%%%%%%%

Stellar mass black holes reside in transient and persistent X-ray
sources. 
Transient X-ray sources transform the gravitational potential energy of the black hole into radiation of the accretion disk, fed by the companion star. From the analysis of the orbital motion in the binary, astronomers obtain information about the gravitational mass, and an estimate of the mass of the black hole. The typical masses of these black holes are around 6-10 Solar mass, while the most massive electromagnetic black holes have masses of
20 $M_{\odot}$
%(Reynolds 2019).
\citep{Reynolds2019}.
Various methods of spin estimates utilize the spectral analysis of radiation from the accretion disk, namely the continuum fitting method or the X-ray reflection spectrum modeling. These methods give consistent results but with a wide range of spin
determinations for individual black holes, from $a=0.3$ to
$a \gtrsim 0.95$.

The newly born black holes are engines of gamma ray bursts. These very energetic events have a transient nature and are associated with a catastrophic collapse of the star. The accretion power is transformed to the bulk kinetic energy of the jet launched along the rotation axis of the black hole. This black hole must be at least moderately or very highly spinning $a \gtrsim 0.6-0.9$ in order to provide an efficient power generation for the jet.

%Supermassive black holes reside in the centers of galaxies.
%All of them capture matter from their cosmic environment in the process of accretion.

Apart from electromagnetic observations, black holes in the Universe are detected via gravitational wave window. The existence of gravitational waves is predicted by General Relativity. The accelerating objects generate changes in the spacetime curvature which propagate outwards with the speed of light. These propagating ripples are called waves, and the observer on Earth will also find the spacetime distorted once such a wave reaches the Solar System. In the gravitational wave detectors, the strain is a measured displacement between the test masses, relative to the reference length. The analysis of the signal is done via numerical relativity methods, and it enables determination of the masses and projected spins
of compact objects whose coalescence is being observed.
%the source of the particular waveform.

Since 2015, the binary compact object mergers, including stellar mass black holes, have been detected many times. These discoveries brought new information about the masses and estimated spins of the black holes produced from stellar progenitors.
In LIGO data, a negative correlation between the black hole masses and the mean effective spins is found \citep{Safarzadeh2020}.
%(Safarzadeh et al. 2020).
In general, the LIGO measurements disfavour large spins. Typical spins are constrained to $a \lesssim 0.4$. For aligned spins, these constraints are tighter, and results suggest $a\sim 0.1$. On the other hand, masses of black holes detected through gravitational waves are systematically larger than previously known. Most of them seem to be around $20-30M_{\odot}$, while the most massive event detected recently was fitted with two black holes weighing about 66 and 85 Solar masses (GW 190521).

In this contribution we are interested in quantifying the gravitational collapse of a massive star and determination of the mass and spin of the newly formed black hole. Our analysis shows that these two quantities are anti-correlated and depend on the angular momentum content in the collapsing envelope.

We build our study following a series of works that have been previously published
%(Janiuk \& Proga 2008; Janiuk et al. 2008; Janiuk et al. 2018; Palit et al. 2019,
\citep{JaniukProga2008, Janiuk2008, Janiuk2018, Murguia2020}.
%Murguia-Berthier et al. 2020).
In particular, \citet{JaniukProga2008} and \citet{Janiuk2008} explored the problem of how fast the black hole can spin up via the collapse. This study addressed the long GRB as a luminous transient powered by the spinning black hole. Depending on the accretion scenario and the angular momentum content in the envelope, the maximum duration of the GRB event can be determined. Basic condition that has to be satified for a successful GRB, is that some part of the rotating envelope must contain enough angular momentum to exceed the critical limit:
\begin{equation}
l_{\rm spec} = l_{0} f(\theta) g(r)
\end{equation}
where the normalization is scaled to $l_{0}/l_{\rm crit}=x$.
%We express $l_{\rm crit}$ as
\begin{equation}
  l_{\rm crit} = {2 G M \over c} \sqrt{2 - A + 2 \sqrt{1 - A}}
  \label{eq:lcrit}
\end{equation}
where $A$ is the black hole dimensionless spin parameter.
The rotating torus and BH spin drive the GRB central engine, as long as the torus angular momentum is above the critical value \citep{Janiuk2008}.
%(Janiuk, Moderski \& Proga 2008).

Various authors
%(e.g. Barkov \& Komissarov 2009??; Lee \& Ramirez-Ruiz???)
\citep{LeeRamirez2006, Barkov2010}
studied the properties of rotating collapsar envelope in the context of long gamma ray bursts.
In particular, also spin-up of the envelope by a companion can be a source of enhanced rotation to prolong the duration and/or provide more power to the transient.
Some results suggest that the binary companion black hole merged with the collapsar's core
might lead to a gravitational wave event accompanied by a bright gamma ray burst \citep{JaniukBejger2017}. On the contrary, other studies show that a certain fraction of massive O-type stars can vanish without a trace (i.e. without a bright luminous transient), if only the slow rotation of those stars prevents them from gaining an effective feedback from accretion disk \citep{Murguia2020}.

\section{The model set-up}

To describe the process of collapse in a proper way, we would need to
start from the matter distribution of an evolved star, and then follow gravitational collapse by solving the Einstein equations for matter-field evolution,
until
the massive Kerr black hole is finally formed and all matter is either accreted or expelled
because of energy deposition, possible due to the shock waves or magnetic reconnections.
Such computations are currently beyond the scope of theoretical and numerical astrophysics.

Some efforts have been made already to simulate a collapsar and involve the conservation equation for the stress–energy tensor. They include the fluid and radiation fields, and the metric evolution followed through the standard BSSN method.
However, the black hole growth was not followed in these works and the simulations stopped after the core collapse
\citep{Ott2018}.
%(Ott et al.2018).
If the black hole was found and diagnosed by means of the baryon mass enclosed inside a certain radius, this radius was identified with the Schwarzschild radius, i.e. the black hole was by definition a non-rotating one.
Its mass is then fixed and also the metric is frozen \citep{Kuroda2018}.
%(Kuroda et al.2018).

In our approach, we focus on the further evolution of the black hole parameters, namely its spin and mass, which are affecting also the Kerr metric changes.
Our approach is therefore more precise than in the above cited works, as for the dynamical evolution studied in General Relativity.
On the other hand, the cost of this approach is a big simplification of the matter field distribution.
We are trying to tackle this problem in two ways.

\subsection{Matter configuration}

First, we adopt the density distribution that is resulting from physical model of the pre-supernova star, pre-calculated by means of the stellar evolution model.
Second, we adopt a  spherical density distribution resulting from the radial accretion problem (i.e. the Bondi solution).
In both cases, we supply the collapsing cloud with a small angular momentum, concentrated on the equatiorial plane, so that the star is rotating. Furthermore, in the Bondi case, we equip the star with magnetic fields of a chosen geometry and strength. 
We study the gravitational
collapse as the sequence of quasi-stationary Kerr solutions for a growing mass and changing spin of the black hole. The spin is changing because of rotating matter is adding the angular momentum after it is transmitted through the black hole horizon.

Computations of the self-similar solutions based on the pre-computed stellar evolution tables
are performed with the numerical code adopted from \citep{JaniukProga2008}.
The magnetized Bondi case is studied by means of the full general relativistic magneto-hydrodynamical simulation, i.e. here the stress-energy tensor contains both matter and electromagnetic parts.
Numerics is tackled here with the generic MHD algorithm adopted from the HARM code
\citep{2003ApJ...589..444G} and further developed by \citep{Janiuk2018}. This code is working in the MPI-parallelized version on the supercomputing clusters and the evolutionary scheme is supplemented with the Kerr metric update embedded in the code, as developed in 2018 by our Warsaw group.

\subsubsection{Density distribution in a pre-supernova star}
We use the pre-supernova models from \cite{Woosley95}
%Woosley \& Weaver (1995),
and also newer ones from \cite{Heger2000, Woosley2002}
%Heger at al. (2000), Woosley et al. (2002),
and \cite{Heger2005}.
%Heger et al. (2005).
The ZAMS mass of the star is 25 $M_{\odot}$.
First two of them did not take into account rotation during the stellar evolution modeling, and neglected magnetic fields.
The initial metallicity was $10^{-4}$, so the mass loss was negligible. The third star model
is magnetized.

\begin{figure}[tbh!]
  \centering
  \includegraphics[width=0.45\textwidth]{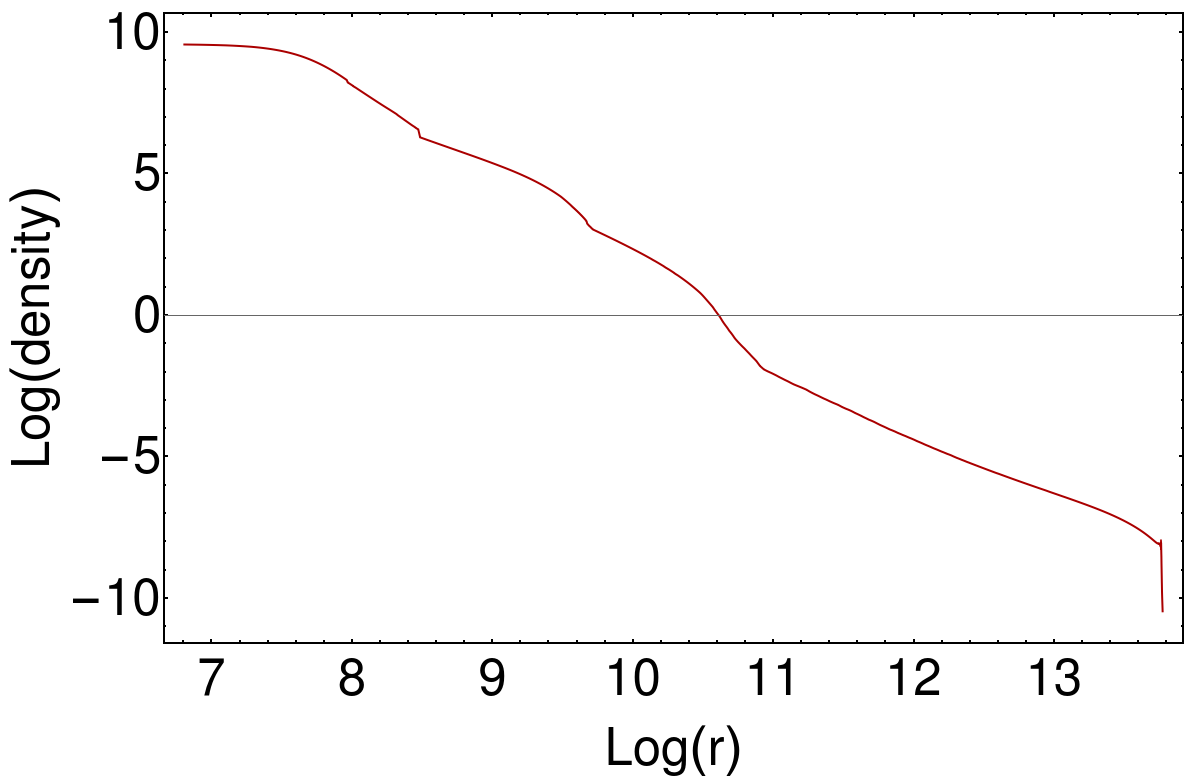}
  \includegraphics[width=0.45\textwidth]{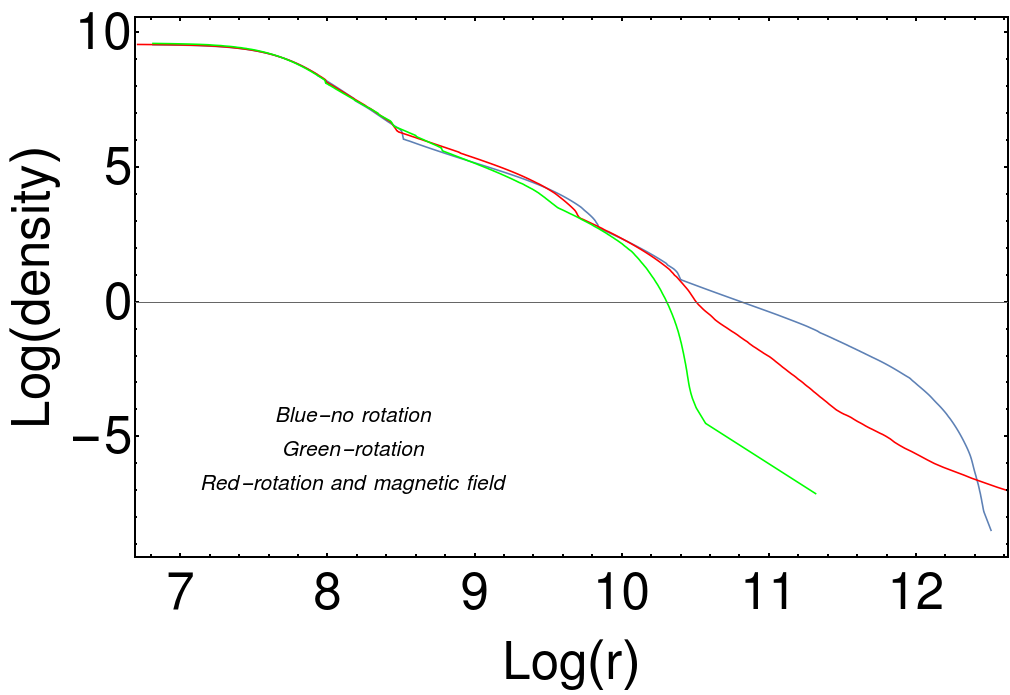}
  \caption{Density distribution in the pre-supernova stars used in our modeling. {\bf Left:} Pre-SN spherically symmetric stellar model from \cite{Woosley95}
    %Woosley \& Weaver (1995),
    taken as initial condition for homologous collapse simulation of \cite{JaniukProga2008}. % Janiuk \& Proga (2008).
   %{\textcolor{red} {Please make the figures with the same box sizes and axes label font sizes}}
    {\bf  Right:} three othe pre-SN stellar models, taken from \cite{Heger2000, Woosley2002}, and
    \cite{Heger2005}. These latest models include rotation or magnetic fields in the stellar evolution calculations.}
 \label{fig_preSN}
\end{figure}

The density distribution in these pre-supernova models is shown in Figure \ref{fig_preSN}.
Subsequent layers of elements synthesized in the stellar interior are traced by this density
profile. The innermost layer, consisting of pure Iron, is forming the core that represents initial
black hole born just at the start of collapse. The mass of this core is equal to 1.4 $M_{\odot}$.
The thin Silikon shell located outside this layer accretes first. Further heavy
shells are then made of
Oxygen with some contribution of Neon, Magnesium and Carbon, and accrete on the newly born black hole. The outermost Helium shell accretes at the core at later times. The Hydrogen envelope, located above radius of $r \gtrsim 10^{11}$ cm, can be either accreted or expelled.

\subsubsection{Spherically symmetric inflow}
We assume initially that the angular momentum of accreted fluid is negligible and its velocity has a non-vanishing component only in the radial direction.
The equation of continuity gives $4\pi r^2\rho(r) u^{r}(r)=-\dot{M}$, where the constant on the right-hand side has a meaning of mass accretion rate.

The distribution of density as the function of radius comes from
the solution of transonic accretion flow in spherical geometry.
The initial density profile and the
radial component of the velocity ($u^r$) of the material is determined by the relativistic version of the Bernoulli equation \citep{Hawley84}.
In this formalism, the critical point ($r_{\rm s}$),
where the flow becomes supersonic, is set as a parameter.
Here we take the value of $r_{s}=80 r_{g}$.
The fluid is considered a polytrope with a pressure $P=K\rho^\gamma$, where $\rho$ is the density,  $\gamma=4/3$ is the adiabatic index, and $K$ is the constant specific entropy.
Once the critical point, $r_{s}$, is set, the velocity at that point is:
\begin{equation}
 (u^r_{\rm s})^2= \frac{GM}{ 2r_{\rm s}}, 
\end{equation}
 where $r$ is the radial coordinate, $M$ is the mass of the BH and the sound speed is:
 \begin{equation}
 \label{eq:cs}
 c_{\rm s}^2=\frac{\gamma\frac{P_{\rm s}}{\rho_{\rm s}}}{1+\frac{\gamma}{\gamma-1}\frac{P_{\rm s}}{\rho_{\rm s}}}.
 \end{equation}
 The constant specific entropy can be obtained using the sound speed:
\begin{equation}
K=\frac{c_{\rm s}^2}{\rho^{\gamma-1}\gamma}
\end{equation}
The radial velocity profile is obtained by numerically solving the equation \citep{Shapiro}:
\begin{equation}
  \bigg{(} 1+\frac{\gamma}{\gamma-1}\frac{P}{\rho}\bigg{)} ^2 \bigg{(}  1-\frac{2GM}{r}+(u^r)^2 \bigg{)} = \rm{constant}
  \label{eq:bernoulli}
\end{equation}

The radial velocity is given by:
\begin{equation}
 (u^r)^2=\frac{GM}{2r}.
\end{equation}

Finally, the  accreting material is 
endowed with small angular momentum scaled to the one at the circularisation radius of
$r_{\rm circ}$, being the ISCO radius (equal to 6 $r_{g}$ for a non-rotating black hole; see \citep{Janiuk2018}).
It is also scaled with polar angle to have its maximum value on the equatorial plane, at $\theta=\pi/2$: 
\begin{equation}
l=S l_{\rm isco}r^2\sin^2{\theta} 
\end{equation}
where $l$ is the specific angular momentum, defined as $l=u^\phi r^2$, $l_{\rm isco}$ is the specific angular momentum at the ISCO of the black hole, and $\theta$ is the polar coordinate. 

\section{Strong gravitational fields}

Gravitational field of a black hole is described by Kerr metric, which can be written in the well-known Boyer-Lindquist coordinate system $(t,r,\theta,\phi)$
%\citep{1973grav.book.....M,1983mtbh.book.....C,1984ucp..book.....W}.
and the metric element is given by:
\begin{eqnarray}
 \textrm{d}s^2 = - \left( 1- \frac{2Mr}{R^2} \right) \textrm{d}t^2 - \frac{4Mra \sin^2\theta}{R^2} \, dt \textrm{d}\phi + \nonumber \\
 + \left( r^2 +a^2 + \frac{2Mra^2}{R^2} \sin^2\theta \right) \sin^2\theta \, \textrm{d} \phi^2 + \frac{R^2}{\Delta} \, \textrm{d} r^2 + R^2\, \textrm{d}\theta^2
% \label{KerrMetric} 
\end{eqnarray}
where $R^2 = r^2 + a^2 \cos^2\theta$, $\Delta = r^2 - 2Mr + a^2$.
The inner horizon is located at $r_{H}=1+(1-a^2)^{1/2} M$, with $a=J/M$,
and  the condition about the presence of the outer event horizon leads to the maximum value of the dimensionless spin, $|a\leq1|$.
Note that here the 
convention is used with $G=c=1$.
Here $M$ denotes mass of the black hole, and the spin parameter $a$ of the Kerr metric describes its rotation.
This spacetime is asymptotically flat and the region far away from the ergosphere and event horizon experiences a negligible gravitational influence.

In Kerr metric the axial symmetry about the rotation axis is assumed
and the metric elements are stationary in time.
The very strong gravitational distorsion becomes infinite and forms a
singularity below the event horizon.

\subsection{Evolution of Kerr metric}

The mass and spin of the black hole are rapidly changing during the collapse process.
In the numerical simulations we update therefore the six non-trivial coefficients of the Kerr matric, according to the change of these quantities.
We neglect however the self-gravity of the accreting fluid, and we assure that the only source of gravitational potential is the dynamically changing black hole mass, $M+\Delta M$:
%$\Delta M = (M^t/M^0-1)$, and $\Delta a = (\dot{J}/M^t - a^{t-dt}/M^{t}\dot{E})\Delta t$
\begin{equation}
\Delta M= \frac {M^t}{M^0}-1
\end{equation}
Also, the black hole spin changes according to the inflow of angular momentum
from the rotating enevelope. Hence,
\begin{equation}
\Delta a = (\frac{\dot{J}}{M^t} - \frac{a^{t-dt}}{M^{t}}\dot{E})\Delta t
\end{equation}
where $M^t$ denotes the current black hole mass at time $t$, $M^0$ denotes initial black hole mass, and $\dot{J}$ and $\dot{E}$ are the flux of angular momentum and energy flux transmitted through the black hole horizon at a given time \citep{Janiuk2018}.
The six non-trivial components of the Kerr metric, namely 
$g_{tt}$, $g_{tr}$, $g_{t\phi}$, $g_{rr}$, $g_{r\phi}$, $g_{\phi\phi}$, are
updated at every time step and they get new values.
This is a simplified treatment of the process, and new simulations with the self-gravity effects taken into account are planned to be the subject of our
future work (Palit et al., in prep.).

\section{Magnetic fields}

In order to initiate the numerical code we employ an initially parabolic magnetic field
%\citep{1974PhRvD..10.1680W,2007IAUS..238..139B},
which is fully described by the only non-vanishing components of the four-potential,
\begin{equation}
%A_{t} &=& Ba\Big[r\Sigma^{-1}\left(1+\cos^2\theta\right) -1\Big], \label{mf1}\\
A_{\phi} \propto (1-\cos(\theta))
%  A_{\phi} &=& B\Big[{\textstyle\frac{1}{2}}\big(r^2+a^2\big)
% -a^2r\Sigma^{-1}\big(1+\cos^2\theta\big)\Big] \sin^2\theta \label{mf2},
\end{equation}
in dimension-less Boyer-Lindquist coordinates.
%and $B$ is the magnetic intensity of the uniform field far from the event horizon.
The magnetic field (and the associated electric component) are generated by currents flowing in the accreted medium far from the black hole, as the latter does not support its own magnetic field. The
%set of two non-vanishing
four-potential vector components define the structure of the electromagnetic tensor, $F_{\mu\nu}\equiv A_{[\mu,\nu]}$; by projecting onto a local observer frame one then obtains the electric and magnetic vectors $\vb*{E}$ and $\vb*{B}$.
From the simple initial configuration, the numerical solution rapidly evolves into a complex entangled structure, with field lines turbulent within the accreting medium.
%and more organized in the empty funnels that develop outside the fluid structure.
At the end of the simulation, while the matter gets accreted into black hole and almost empty envelope remains, the field lines become more organized again.

\section{Results}

\subsection{Homologous mass accretion in collapsar}

\begin{figure}[tbh!]
 \centering
  \includegraphics[width=0.45\textwidth]{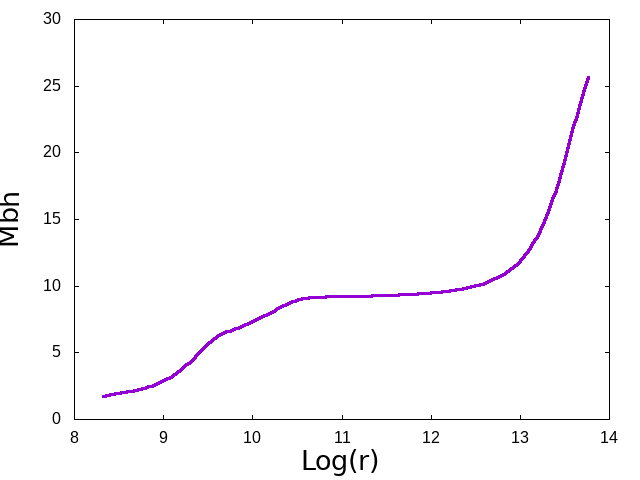}
 \includegraphics[width=0.45\textwidth]{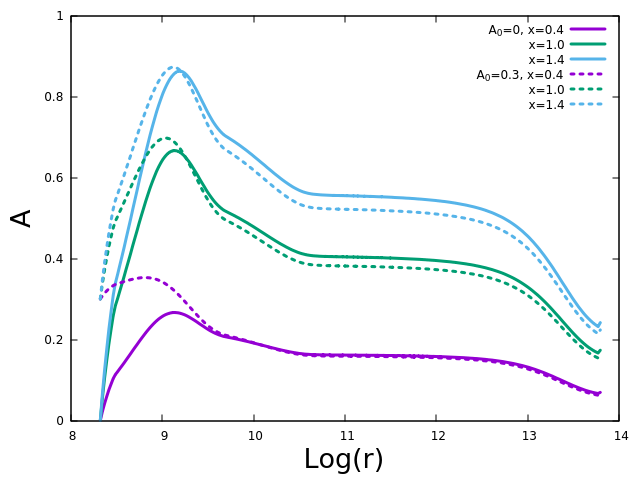}
 \caption{BH mass (left) and spin evolution (right) during collapse. Initial density distribution is taken from the  25 $M_{\odot}$ pre-supernova model of \cite{Woosley95}.
   %Woosley \& Weaver (1995).
   Accretion scenario assumes homologous collapse with black hole spin-up, starting from either spin-less or moderately spinning core ($A_{0}=0$ or $A_{0}=0.3$). Initial core mass is $1.4 M_{\odot}$. Angular momentum in the envelope is maximum at the equator, and scales with polar angle as $\sin^{2}\theta$, where $\theta=0$ and $\theta=\pi$ refer to the poles. Normalisation of rotation with respect to the critical value, see Eq.\ref{eq:lcrit}, is denoted with $x=0.4, 1.0$ and 1.4 (see blue, green and cyan lines in the right panel). }
 \label{fig_mathsim}
\end{figure}

%{\bf DESCRIBE the process in few senetencess here}

The original paper by \cite{JaniukProga2008}  included four models of the angular momentum profile, but did not consider the black hole spin changes during the collapse. Then the subsequent work \cite{Janiuk2008} examined the black hole spin evolution, with the number of models of angular momentum distribution limited to two cases. Here we present the results of calculations for one of the  models which was not included in the second paper, with the angular momentum profile described by the function: 
\begin{equation}
	f(\theta) \propto \sin^{2}(\theta).
\end{equation}
with several normalizations with respect to the critical vaue, $l_{crit}$ (see Eq.\ref{eq:lcrit}): $x=0.4$, $x=1.0$ and $x=1.4$. Note that the same values are used in GRMHD simulations presented in te next Section.

In this particular model we allowed for accretion of matter with super- and sub-critical angular momentum at the same time, thus referring to a homologus accretion scenario. We also did not terminate our calculations after there was no matter with sufficient amount of angular momentum to sustain the torus, so that no part of the Hydrogen envelope was expelled, and finally all the mass is accreted. We performed the calculations for two values of the initial black hole spin: $A_0=0$ and $A_0=0.3$. The spin and the black hole mass evolution are shown in Fig. \ref{fig_mathsim}.
The black hole mass increase is the same for all the $x$ normalization values, because matter with sub- and super critical angular momentum accrete togethter. The black hole spin evolution depends on both initial spin and initial angular momentum of the matter. In case of $x=0.4$ maximal spin is significantly higher for $A_0=0.3$ than for a non-spinning black hole. The difference between maximal spin values in the models with $A_0=0.3$ and $A_0=0$ is smaller for higher $x$. In general, we note that the spin starts to increase immediately at the beginning of the calculations, and the lower $x$ value is, the faster the spin reaches its maximum and then starts to decrease.
The final spin value does not depend on the $A_0$.

\subsection{MHD evolution of slowly-rotating inflow with Kerr metric update}

In the $HARM$ code GR MHD simulations, we set the outer boundary of the computational domain at the radius $10^3r_{\rm g}$, where the inflow is purely radial at initial time. The inner boundary is set at $\simeq 0.98r_{\rm H}$, i.e.
at below the horizon radius for the corresponding value of spin $a$.
%{\bf CHECK the inner radius value?}
The grid domain has been resolved at $256 \times 256$ points in $(r,\theta)$ coordinates.
%and the polytropic index set to $k=4/3$.
In MHD models, we renormalize the magnetic intensity to determine the plasma parameter $\beta=p_{\rm gas}/p_{\rm mag}$ (smaller $\beta$ corresponds to a more magnetized plasma). 

%In order to initialize the computation we employ the hydrodynamic (non-mag\-net\-ized), purely spherical inflow. We set $\beta\rightarrow\infty$ and allow the inflow to build a steady-state Bondi accretion at a certain level of $\dot{M}$ (see fig.~\ref{fig1}). Once the inflow stabilizes to a quasi-steady state inflow, we impose the large-scale Wald magnetic field along the rotation axis, which is then evolved further.
Because of the perfect conductivity and the force-free approximation (apart from the effective small-scale numerical dissipation), the magnetic field lines remain attached to plasma.
%However, the evolution of the system can be strongly altered if the magnetic field is strong enough, so that its repulsive tendency halts accretion.
%This effect is governed by the magnetization
During the evolution, the $\beta$-parameter is not uniform across the computational domain and it changes in time.
In the limit of negligible magnetization ($\beta = \infty$) the gravitational attraction of the black hole prevails. But in the case of equipartition between the magnetic and hydrodynamic pressure ($\beta\approx1$) near the horizon  the accretion rate is diminished
%(see fig.~\ref{fig2}).
(cf. \cite{Karas2020}, \textit{this Proceeding}).
%Let us note that the mass and spin of the black hole are not updated during the simulation because the amount of accreted material is tiny compared to the black hole mass.

%\subsection{General relativistic hydrodynamical simulations}

%{\bf Describe the evolution of non-magnetized flow in few sentences. Pick several values of final black hole mass and spin, for given values of initial spin and envelope rotation.   Mention the mass scale and density scale, and the temperature range, as resulting from sonic radius parameter.}

\begin{figure}[tbh!]
  \centering
  \includegraphics[width=0.32\textwidth]{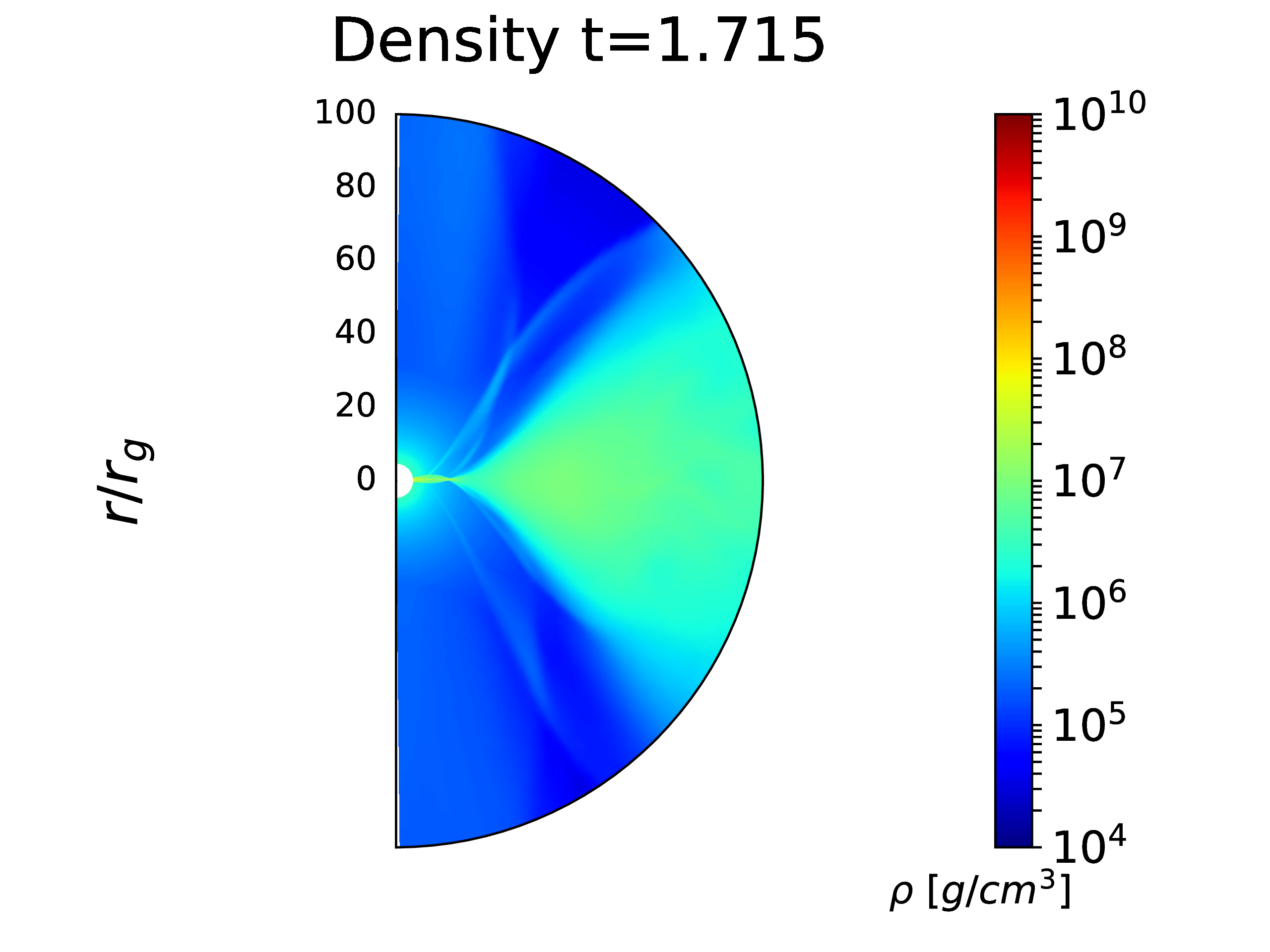}
  \includegraphics[width=0.32\textwidth]{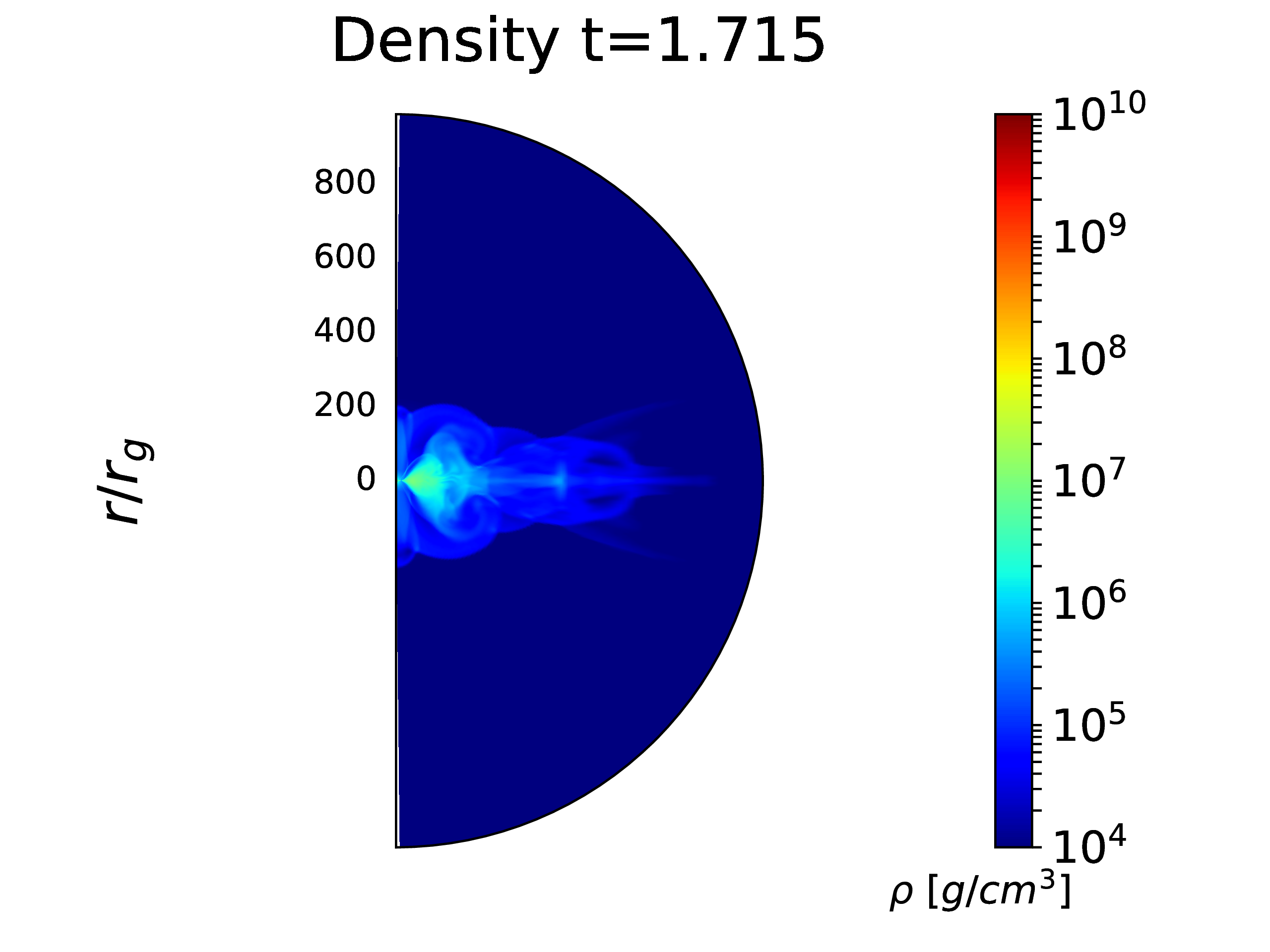}
  \includegraphics[width=0.32\textwidth]{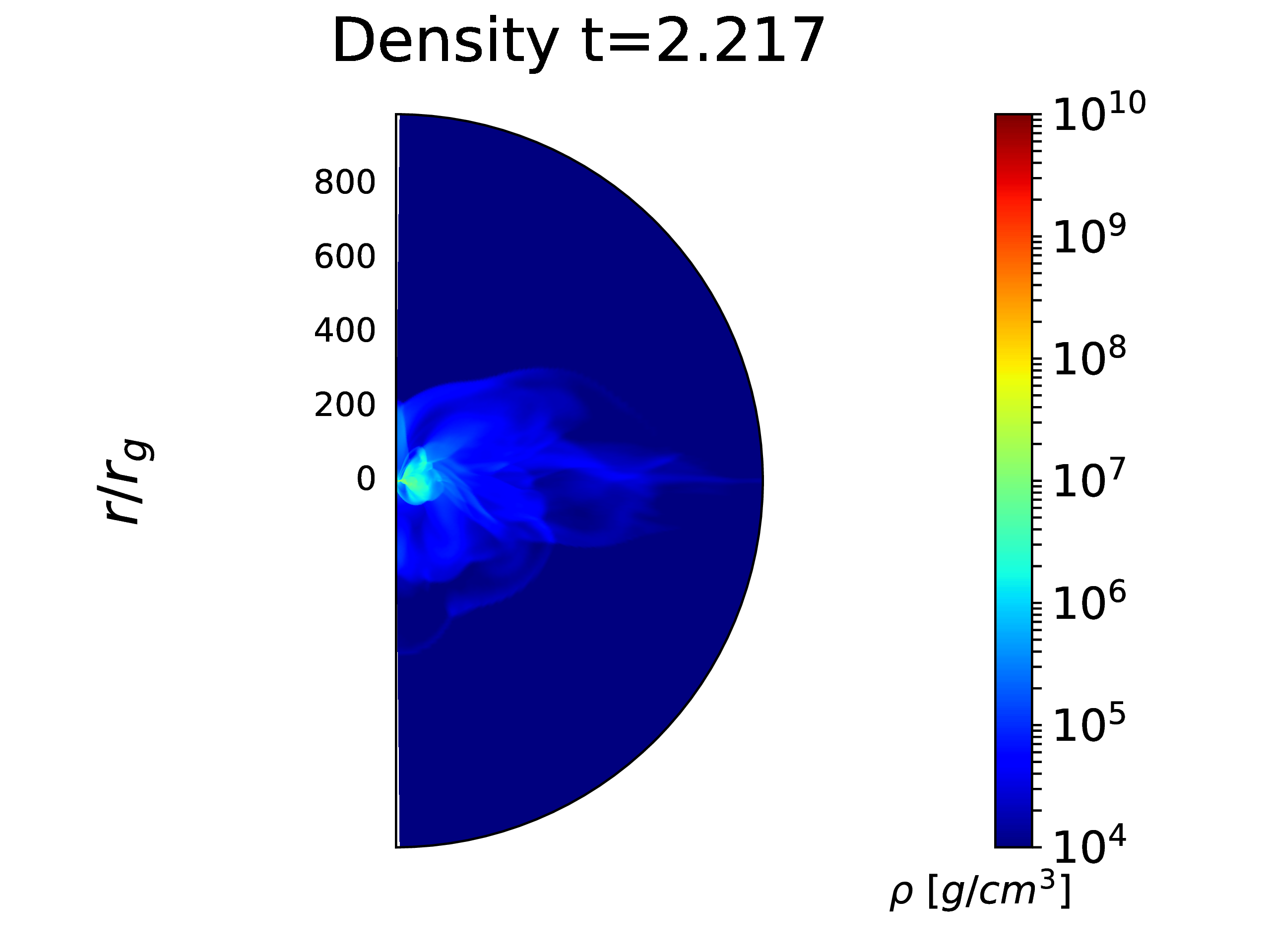}
 \caption{Density distribution in the late time of the simulation. Model parameters: black hole initial spin $A_{0}=0.5$, critical rotation parameter, $S=1.0$. Model neglects magnetic fields. Rotationally supported disk-like structure is present for a long time in the equatorial plane. Note the different spatial scale of the left panel, which shows a zoom-in of the middle figure.
   %{\bf \textcolor{red}{Pick some characteristic snapshots of density profiles for a hydro model,which were not presented in ApJ paper}}
   }
 \label{fig_A05S10}
\end{figure}

Our collapsar simulations start from a spherically symmetric distribution of density of a purely radial infall, which is quickly broken by the imposed roation. The flow concentrates towards the equator, forming a mini-disk structure.
The flow is supersonic near the black hole horizon, and in many cases, the multiple sonic surfaces are found with an aspherical shape (resembing an \textit{eight-letter}). This feature refers to the inner sonic surface which after some time gets accreted. In the models with critical and supercritical roation ($S \gtrsim 1.0$),
this initial transient shock is accompanied with some moderate variability of the accretion rate. 
Another sonic surphace, located initially at 80 $r_{g}$, expands outwards. The expanding shock velocity is typically much lower than the escape velocity at the shock radius.

The densest part of a roationally-supported mini-disk is enclosed within a
small region of $r< 20 r_{g}$ (cf. \cite{Murguia2020}).
The sub-critical models ($S\le1.0$) do not contain enough angular momentum to form a mini-disk bubble, and the material from both polar regions and equator
can contribute to accretion and black hole mass grows more quickly in these models. In case of super-critical accretion, only material from polar regions accretes, while the angular momentum cannot be transported if magnetic fields are neglected. The accretion rate in this case is rather low for the first part of the simulation, while it grows later, when the bulk of material falling from the outer parts of the envelope reaches the mini-disk and is able to overpass it
above and beow the equatorial plane. This phase (reached typicaly after $t>0.8-1$ s, in physical time units) is also associated with large spikes in the accretion rate. The mini-disk is destroyed, nevertheless in many simulations we observe the existence of a long-living disk-like structure a the equatorial plane, which is sustained until the end of the simulations (typically $t_{end}=2$ s). The detailed shape, and time for which the feature is preserved, depends also somewhat on the value of initial back hole spin $A_{0}$ (for the highets probed value, $A_{0}=0.85$, we found the longest timescale of disk structure, $t_{f}=4.5$ s; see \citep{Krol2020}). 
In Figure \ref{fig_A05S10} we show the density distribution in the late phase of the simulation, for one exemplary model. Parameters of this model are $A_{0}=0.5$ and $S=1.0$. 

The evolution of the black hole spin is non-linear, as the rotation of the
black hole can both speed up and slow down, depending on the amount of
angular momentum that is reaching the horizon. The maximum value of the spin reached during the collapse temporarily, as well as the final value, depends also on the assumed initial spin.
For sub-critical rotation models, only for smallest value of $A_{0}=0.3$
we observed a temporal spin-up of the black hole (up to $A\sim0.4$), while the end value was below the starting one ($A_{f}\sim0.15$).
For super-critical rotation, the black hole could even spin-up maximally for some period of time, but finally the spin was smaller. Typically $A\sim0.7$ was reached in case of angular momentum in the envelope normalised to $S=1.4$, which means that effectively the highly spinning black hole at $A_{0}=0.85$ did spin-down after the collapse.

The final black hole masses are oboviously limited  by the total mass of he envelope, assumed always to be 25 $M_{\odot}$. They were between $M_{BH}^{end} = 11$ and 18 $M_{\odot}$ at the end of the simulation, and in non-magnetized models the largest back hole masses were obtained for sub-critical rotations, which also correlates with the smallest final spins. These values did not differ much between the models with various initial spins.
For super-critical rotation of the envelope, the final black hole mass was smaller, and also decreased for large initial black hole spins.

%\subsection{Magneto-hydrodynamical model of slowly-rotating quasi-spherical flow in GR}

\begin{figure}[tbh!]
 \centering
 \includegraphics[width=0.32\textwidth]{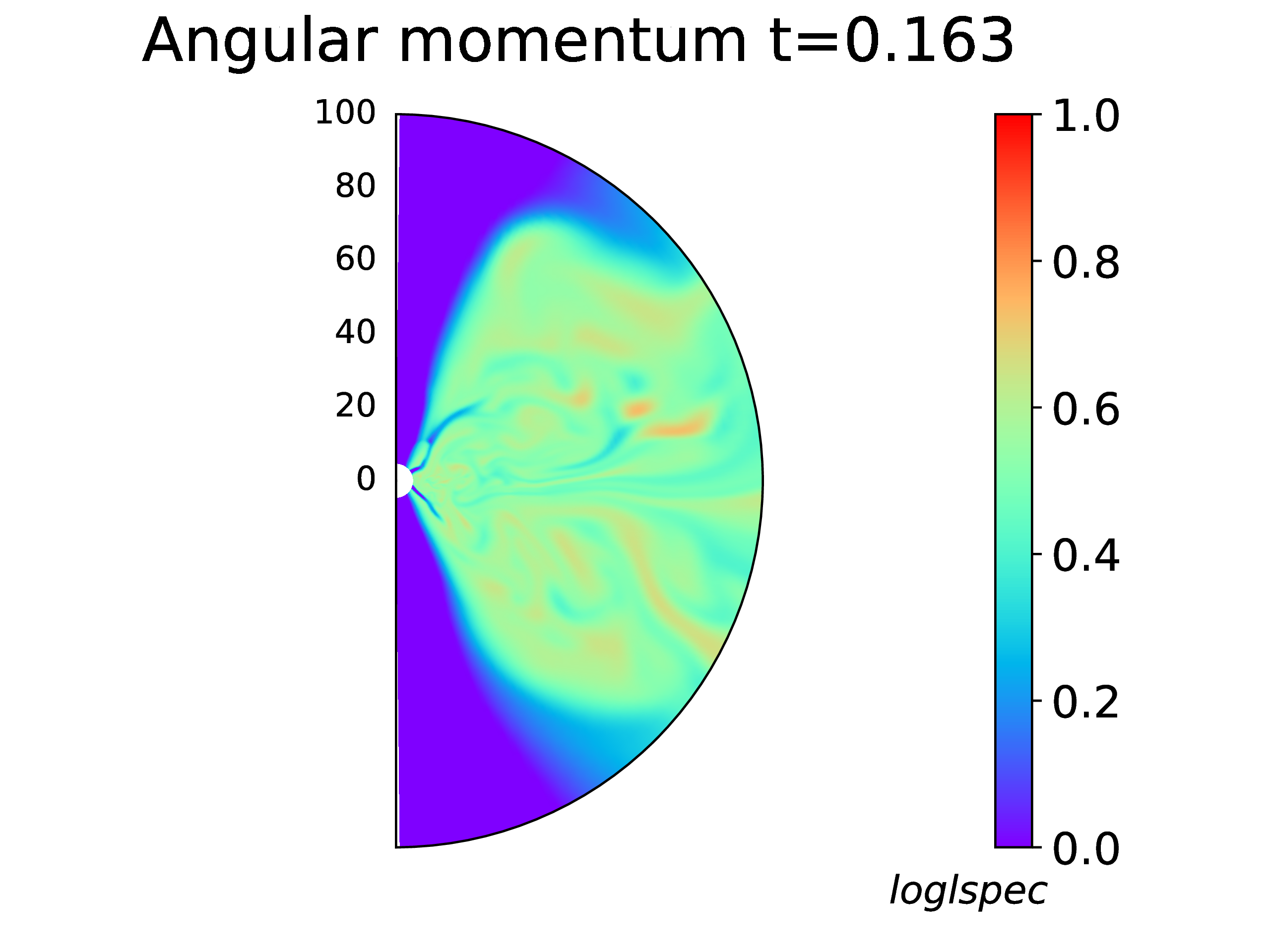}
 \includegraphics[width=0.32\textwidth]{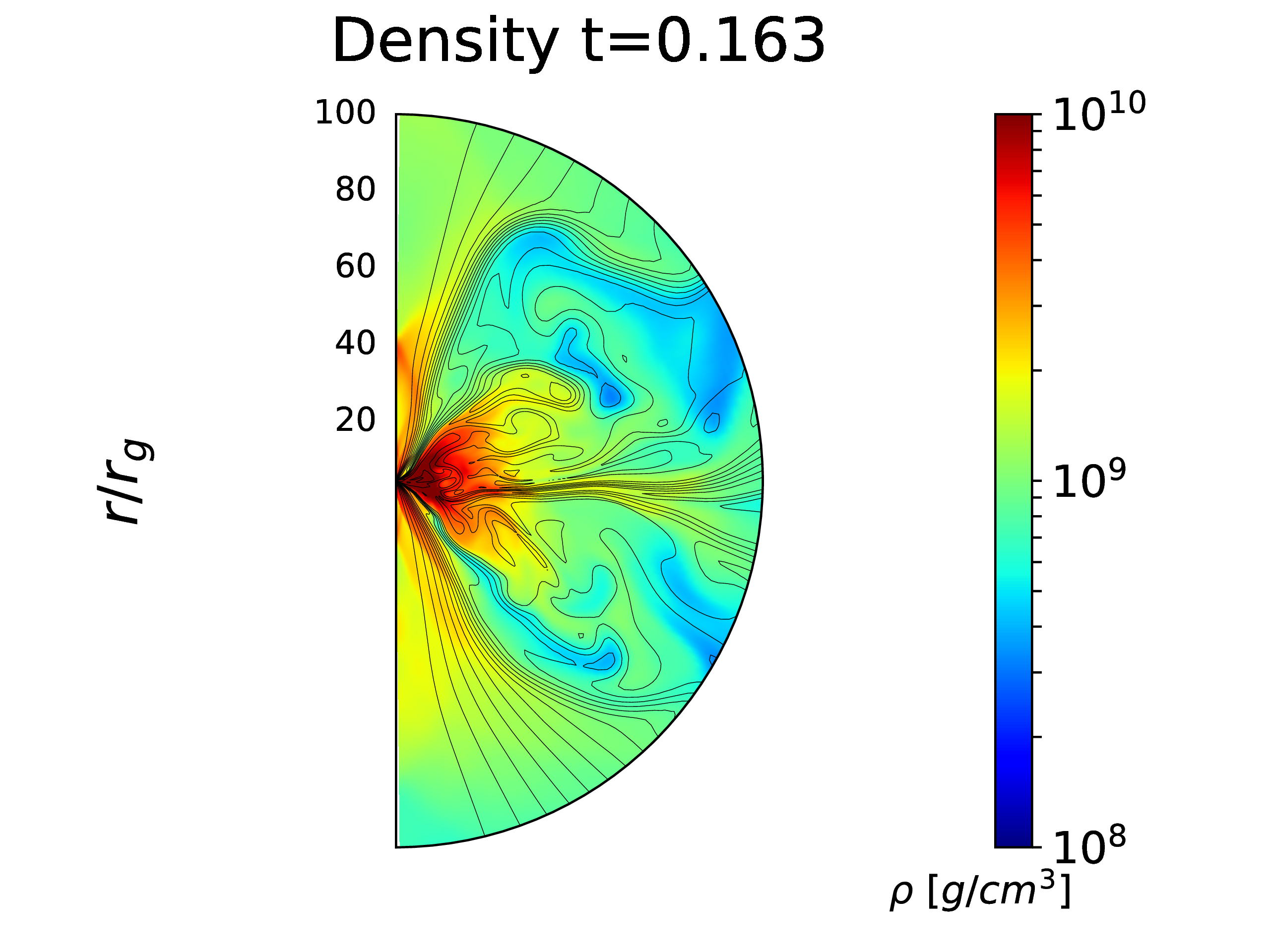}
 \includegraphics[width=0.32\textwidth]{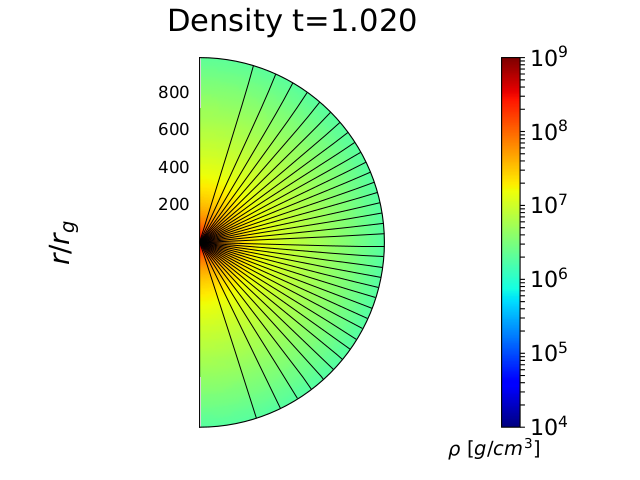}
 \caption{Results for magnetized model, with $\beta=1$, initial black hole spin $A_{0}=0.5$ and critical rotation parameter, $S=1.0$. Figures show angular momentum distribution and density field with overplotted magnetic field lines at the beginning of the simulation (left and middle panels) and the density structure, at the end of the simulation. Note the different spatial scale of the right panel, which shows a zoom-out of the middle figure.}
 \label{fig_A05S10beta1}
\end{figure}

In order to reveal the changes of the magnetic field near the black hole, we study the evolution of the strongly  magnetized plasma, that is inflowing into the horizon.
In Figure \ref{fig_A05S10beta1} we present the MHD simulation results. Parameters of rotation in the envelope and initial black hole spin are the same as in
Fig. \ref{fig_A05S10}, but now the model is magnetized, and the magnetic to gas pressure ratio in the accreting cloud is equal to $\beta=1$.

\begin{figure}[tbh!]
 \centering
 \includegraphics[width=0.32\textwidth]{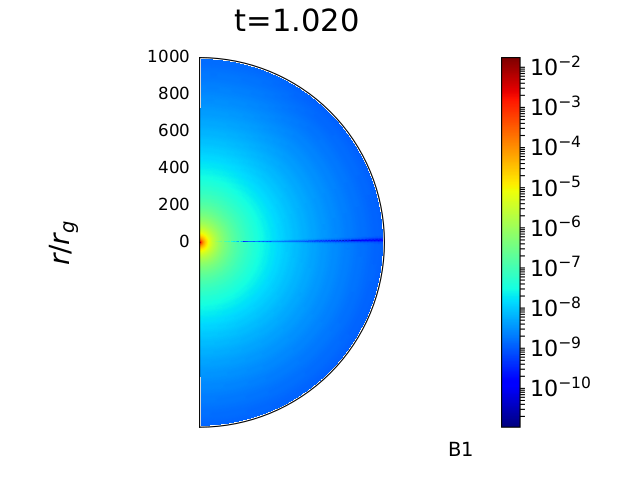}
 \includegraphics[width=0.32\textwidth]{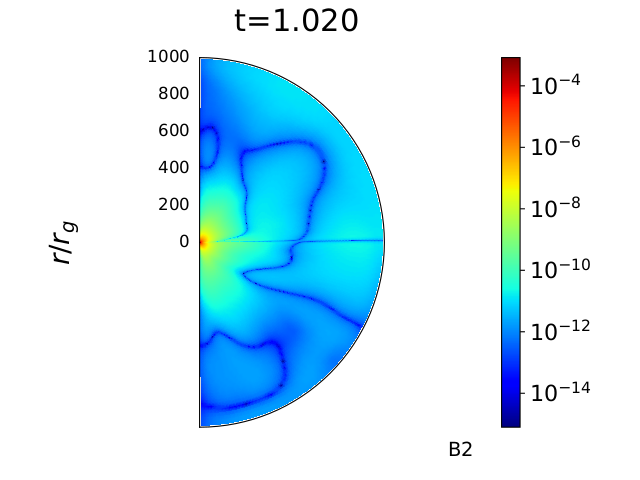}
 \includegraphics[width=0.32\textwidth]{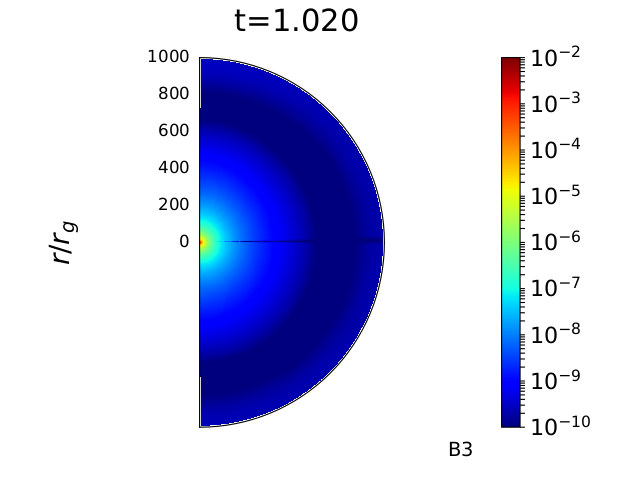}
 \caption{Components of magnetic field 3-vector, at the late stage of the simulation. Parameters of the run: $\beta=1$, initial black hole spin $A_{0}=0.5$, envelope rotation  $S=1.0$}
 \label{fig_Bcomp}
\end{figure}

As mentioned above, the initial configuration is the parabolic magnetic field solution, but this configuration starts quickly changing once the inflowing plasma arrives in the domain, while the magnetic field is coupled to matter.
The purely poloidal field changes and develops a strong toroidal component
(see Fig. \ref{fig_Bcomp}).
%\footnote{Some more complicated configurations were explored by \citet{2020arXiv200407535K}. Interestingly, these authors found that a parabolic magnetic field also develops in accretion torus funnel around the vertical axis, for any initial magnetic field configuration.}
%As was also mentioned above, the initial configuration for the plasma inflow is the Bondi solution. As soon as a field line enters the ergosphere, it has to terminate at the event horizon. Therefore, the magnetic field lines in a force-free magnetosphere are not expelled by even extreme rotation of the black hole.
%Figure~\ref{fig4} shows the magnetic flux as a function of time and fig.~\ref{fig5} exhibits a more detailed view of the brief initial period.

%Finally we vary the value of the Wald magnetic intensity $B$ parameter to obtain more initially magnetized (lower plasma $\beta$).\footnote{Notice that the code itself poses a limit $\beta>10^{-4}$ under which it starts introducing artificial density floor in order to avoid numerical integration problems.} Figures \ref{fig6}--\ref{fig7} show several snapshots of the magnetic field and plasma density at different resolution. An equatorial outflow forms at late stages of the system evolution.

%{\bf{ADD Figures with accretion rate and bh spin growth, for A0=0.5, S=1.0. THREE models, non-magnetized, and magnetized with beta=1, and beta=100, on same plots.}}

%{\bf Decribe the evolution of magnetized flow in several sentences. Mention the resulting changes in the black hole spin and mass.   Mention beta vs. Alvfen to sound speed ratio.}

We observe also that the magnetized jet wants to form in the polar regions, where the
open field lines are visible in the early stage of evolved configuration, together with dense and turbulent torus structure in the equatorial plane (time $t\sim 0.1-0.15$ s). Nevertheless, because of large density in the envelope, the jet cannot break out of the collapsing star.
The polar funnels are baryon-polluted, and quickly change the magnetic field configuration back to radially-dominated, and the equatorial configuration of matter turns back to quasi-spherical (at time $t\sim 0.5$ s).
We envisage, that full 3-dimensional simulations might be needed to overcome this problem and allow to sustain a long-living, dense and magnetized torus together with a
jet-like funnel.

\begin{figure}[tbh!]
 \centering
 \includegraphics[width=0.32\textwidth]{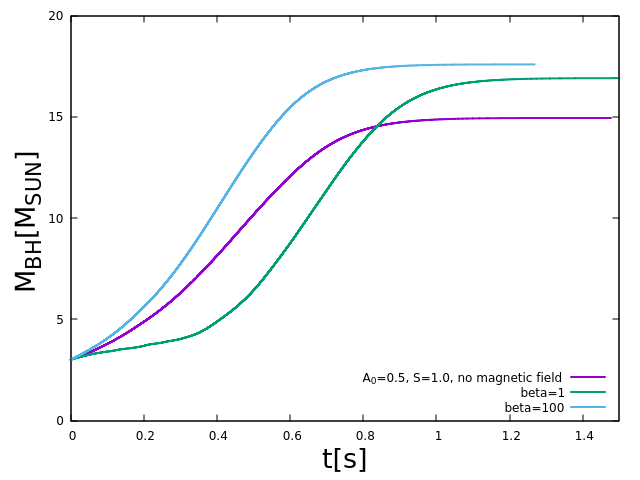}
 \includegraphics[width=0.32\textwidth]{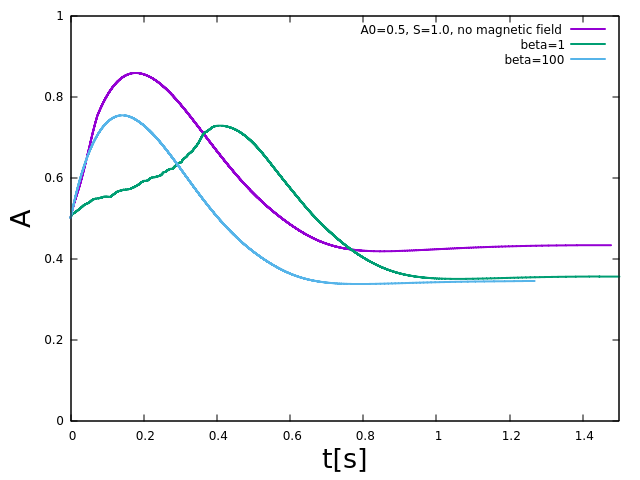}
 \includegraphics[width=0.32\textwidth]{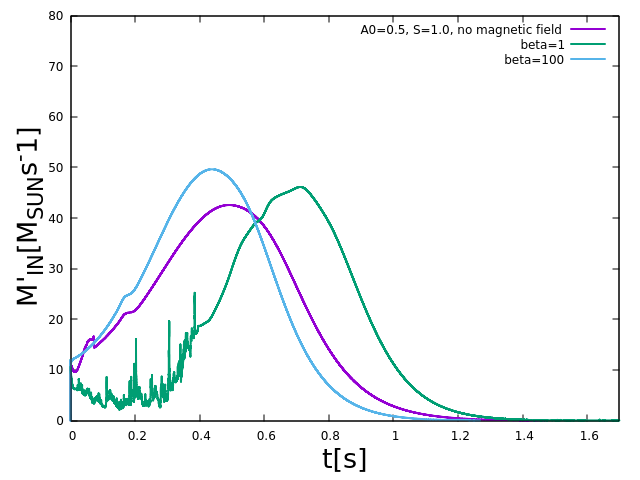}
 \caption{Evolution of the black hole mass, its spin, and accretion rate onto BH horizon, during the simulation. Parameters of the run: initial black hole spin $A_{0}=0.5$, envelope rotation parameter $S=1.0$. Non-magnetized models are shown with purple lines. Magnetic field in magnetized model was parameterized with $\beta=1$ (green lines) or $\beta=100$ (cyan lines).}
 \label{fig_accrate}
\end{figure}

We also probed the effects of magnetic fields on the accretion rate, black hole spin, and mass in the collapsar models.
In Figure \ref{fig_accrate} we show the evolution of these quantites, for initial spin
$A_{0}=0.5$ and the critical rotation parameter $S=1.0$.
Parabolic magnetic field was normalized to $\beta=1$ or $\beta=100$. For comparison, non-magnetized model of the same parameters but $\beta=\infty$ is presented in the Figure.
%\bf{ADD Figures with beta1/beta100}
The main quantitative difference is that the black hole spin is always smaller at its maximum in the magnetized models, than in the non-magnetized, for the same set of other parameters.
The final black hole spin is also smaller if magnetic fields are included.
On the other hand, the mass of the black hole achieved larger values. This is the result of angular momentum transport via magnetic fields, which allows matter to accrete not only from the polar regions, but also through the rotating disk.
The variability of accretion rate is more visible in the weakly magnetized case ($\beta=100$) than in the strongly magnetized. We suggest that this is an effect of a very strong magnetic barrier in the latter.

\section{Conclusions}
\begin{itemize}
\item We compute the collapsar model with slowly-rotating quasi spherical collapse with changing black hole spin and mass and Kerr metric update. We probed a range of angular momentum contents in the collapsars envelope, and range of initial black hole spins.
\item Our method to follow collapse is fully GR MHD, while still not by exactly solving the Einsteins equations, but it gives a good approximation to this problem
%some attempts (ET and microphysics) have been made recently (Kuroda et al. 2018) but with Schwarzschild metric.
\item Our test models out some constraints on the angular momentum content of the collapsing progenitor star, with the resultant mass and spin of the BH.
  \item For supercritical rotation, we always observe spin up of the black hole at some stage of the simulation. The dependence on the initial black hole spin is however not monotonic, and at the end of simulation, the black hole can be effectively spun down with respect to the initial spin value.
  \item The strongly magnetized collapsars reach lower maximum black hole spins, and
    even for supercritical rotation in the envelope, the black hole may not reach the maximum Kerr parameter.
    \item The growth of the black hole mass is largest when the envelope rotation is slow, and when the black hole was at least moderately spinning initially.
\item Two shock fronts were found with velocities are 0.014 c and 0.022 c.  For the models with sub critical envelope rotation we obtained higher velocities of  the  shock  fronts:  0.04c and  0.044c, depending on BH spin ($A_{0}=0.5, 0.85$). This  trend  seems  opposite  in comparison to
  \cite{Murguia2020}
  %Murguia-Berthier et al. (2020),
  who studied non-spinning BHs.
%\item Further work: the initial conditions as results from the state-of-the-art stellar evolutionary model, and constraining qualitatively the feedback limits. Three dimensional modeling.
\end{itemize}

\ack%%%%%%%%%%%%%%%%%%%%%%%%%%%%%%%%%%%%%%%
The authors thank Daniel Proga for helpful discussion.
AJ was supported by grant no. 2019/35/B/ST9/04000  from  the Polish National Science Center, and acknowledges computational resources of the Warsaw ICM through grant Gb79-9. D.K. was supported by Polish NSC grant 2016/22/E/ST9/00061.
%{\bf Other???}


\begin{thebibliography}{99}

%%%%%%%%%%%%%%%%%%%%%%%%%%%%%%%%%%%%%%%%%
%\bibliographystyle{ragtime} % style aa.bst
%\bibliography{zajacek} % bib file
\bibitem[Barkov \& Komissarov(2010)]{Barkov2010} Barkov, M., \& Komissarov, S. 2010, MNRAS,
  401, 1644
\bibitem[Gammie et al.(2003)]{2003ApJ...589..444G} Gammie, C.~F., McKinney, J.~C., \& T{\'o}th, G.\ 2003, ApJ, 589, 444

\bibitem[Hawley et al. (1984)]{Hawley84} Hawley, J.~F., Smarr, L.~L. \& Wilson, J.~R. 1984, ApJ, 277, 296
  
\bibitem[Heger et al. (2000)]{Heger2000} Heger, A., Langer, N., \& Woosley, S. E. 2000, ApJ, 528, 368
  
\bibitem[Heger et al. (2005)]{Heger2005} Heger, A., Woosley, S. E., \& Spruit, H. C.\ 2005, ApJ, 626, 350

\bibitem[Janiuk \& Proga (2008)]{JaniukProga2008} Janiuk, A., \& Proga, D. 2008, ApJ, 675, 519

\bibitem[Janiuk et al. (2008)]{Janiuk2008} Janiuk, A., Moderski, R., \& Proga D. 2008, ApJ, 687, 433
  
\bibitem[Janiuk et al. (2017)]{JaniukBejger2017} Janiuk, A.,  Bejger, M., Charzynski, S., Sukova, P. 2017, New Astronomy, 51, 7
  
\bibitem[Janiuk et al. (2018)]{Janiuk2018} Janiuk, A., Sukova, P., \& Palit, I. 2018, ApJ, 868, 68
  
\bibitem[Karas et al. (2020)]{Karas2020} Karas, V., Sapountzis, K., Janiuk, A. 2020, in RAGtime 20-22: Workshops on Black Holes and Neutron
Stars. (in press)

\bibitem[Kr{\'o}l \& Janiuk (2020)]{Krol2020} Kr{\'o}l, D., \& Janiuk, A. 2020, ApJ (submited)  
  
\bibitem[Kuroda et al. (2018)]{Kuroda2018} Kuroda, T., Kotake, K., Takiwaki, T., Thielemann, F.-K. 2018, MNRAS, 477, L80 

\bibitem[Lee \& Ramirez-Ruiz (2006)]{LeeRamirez2006} Lee, W.H., \& Ramirez-Ruiz, E. 2006, ApJ, 
  641, 961
  
\bibitem[Murguia-Berthier et al. (2020)]{Murguia2020}  Murguia-Berthier, A., Batta, A., Janiuk, A., et al. 2020, ApJL, 901, 24

  \bibitem[Ott et al. (2018)]{Ott2018} Ott, C.~D., Roberts, L.~F.; da Silva Schneider, A., Fedrow, J.~M., Haas, R., Schnetter, E. 2018, ApJL,   855, 3
  
\bibitem[Reynolds (2019)]{Reynolds2019} Reynolds, C.S. 2019, Nature Astronomy, 3, 41
 
\bibitem[Safarzadeh et al. (2020)]{Safarzadeh2020} Safarzadeh, M., Farr, W. M., \& Ramirez-Ruiz, E. 2020, ApJ, 894, 129

  \bibitem[Shapiro \& Teukolsky (1986)]{Shapiro} Shapiro, S.L., \& Teukolsky, S.A., 1986, ``Black Holes, White Dwarfs and Neutron Stars: The Physics of Compact Objects''
  
\bibitem[Woosley \& Weaver (1995)]{Woosley95} Woosley, S. E. \& Weaver, T. A. 1995, ApJS, 101, 181
  
\bibitem[Woosley et al. (2002)]{Woosley2002} Woosley, S. E., Heger, A., \& Weaver, T.A. 2002, Rev. Mod. Phys., 74, 1015
  
  \end{thebibliography}
\end{document}